\documentclass[fleqn,10pt]{wlscirep}
\usepackage[utf8]{inputenc}
\usepackage[T1]{fontenc}
\usepackage{xcolor}
\title{Natural experiments from {Earth Hour} reveal urban night sky being drastically lit up by few decorative buildings} 

\author[1,*]{Chu Wing So}
\author[1]{Chun Shing Jason Pun}
\author[1,2]{Shengjie Liu}
\author[3,4,5]{Sze Leung Cheung}
\author[5]{Ho Keung Kenneith Hui}
\author[4]{Kelly Blumenthal}
\author[6]{Constance Elaine Walker}
\affil[1]{Department of Physics, The University of Hong Kong, Pokfulam, Hong Kong, China}
\affil[2]{Spatial Sciences Institute, University of Southern California, Los Angeles, CA 90089, USA}
\affil[3]{Faculty of Science, The University of Hong Kong, Pokfulam, Hong Kong, China}
\affil[4]{Office for Astronomy Outreach, International Astronomy Union, National Astronomical Observatory of Japan, Mitaka, Japan}
\affil[5]{Ho Koon Nature Education cum Astronomical Centre (Sponsored by Sik Sik Yuen), 101 Route Twisk, Tsuen Wan, New Territories, Hong Kong, China}
\affil[6]{U.S. National Science Foundation National Optical-Infrared Astronomy Research Laboratory, 950 N. Cherry Ave., Tucson, AZ 85719, USA}
\affil[*]{socw@connect.hku.hk}

\begin{abstract}
Light pollution, a typically underrecognized environmental issue, has gained attention in recent years. While controlling light pollution requires sustained efforts, Earth Hour offers a unique natural experimental setting to assess temporary lights-out measures. Using photometric and spectroscopic sensors, we observed up to 50\% night sky darkening during Earth Hour from 2011 to 2024 in Hong Kong, primarily as a result of a small but critical number of lights-out instances in central business districts, as evidenced by crowd-sourced photography records. Weekend lighting pattern in the city during Earth Hour remained unaffected. The emission reductions mostly occurred in the 445–500, 500–540, and 615–650 nm spectral ranges\textemdash corresponding to peak emissions from LED billboard screens\textemdash and in the 585–595 nm range, associated with metal halide floodlights used for facades and billboards. Our study identifies these sources as major contributors to urban light pollution. By combining multimodal observations, we offer a comprehensive assessment of light pollution sources and the potential benefits of sustainable lighting practices in urban environments. This research highlights the importance of targeted light pollution mitigation efforts and provides critical insights for policymakers to enhance urban sustainability and human well-being.
\end{abstract}
\begin{document}

\flushbottom
\maketitle

\thispagestyle{empty}

\section*{Introduction}
Light pollution, the improper use and abuse of artificial light at night (ALAN), has emerged as a significant global environmental issue with wide-ranging impacts~\cite{arroyo:2024,fabio:2016}. Research has highlighted how ALAN can disrupt natural day-night cycles, affecting physiology, behavior and evolution in flora and fauna\cite{IDA:2022,jones:2015,abraham:2013}. Excessive or variable light near human habitats has been associated with suppressed melatonin production, potentially affecting human health and circadian rhythms~\cite{blume:2019,jones:2015}. Furthermore, ALAN-induced skyglow reduces the quality and precision of astronomical observations~\cite{falchi:2020b}. These diverse effects emphasize the importance of understanding and mitigating the environmental impact of light pollution through comprehensive and interdisciplinary approaches~\cite{comino:2023,kyba:2020a}.

External ALAN that contributes to light pollution serves various functions. It is often designed to illuminate streets and public areas to enhance safety and security, remaining operational throughout the night, while also being used for decorative and advertising purposes, operating according to varied business schedules.
Given the diverse geographical and temporal practices surrounding external ALAN usage \cite{pun:2014,pun:2012}, accurately assessing ALAN's impacts on the urban environment poses significant challenges. 
In rare instances where lighting inventories are available \cite{gokus:2023,barentine:2020,levin:2025}, their completeness, such as geographical coverage and inclusion of privately owned fixtures or not, along with their reliability and timeliness, cannot be guaranteed. Operational details, including schedules, brightness and angle of illumination\textemdash key factors contributing to skyglow\textemdash are often absent from these inventories, which typically provide only basic statistics.
Recent advances include manual night surveys to document the various types of ALAN perceived by pedestrians \cite{gokus:2023,nachtlichter:2025}. However, without dedicated coordination, sufficient manpower and resources such as mobile applications, replicating these street surveys in additional cities remains a challenging endeavor.

The physical impacts of ALAN on light pollution are immediate, suggesting that one way to assess their overall contribution to the environment is to turn them off, either unintentionally or intentionally, and observe the differences. Large-scale power outages caused by electrical infrastructure incidents, natural disasters \cite{miguel:2019,levin:2023b,mo:2025} or human conflicts \cite{li:2013a,levin:2018} offer accidental opportunities for such assessments. However, since nearly all kinds of lighting (except for a minority powered by batteries or emergency generators) are likely extinguished during power losses, this approach often fails to distinguish the contribution of different light sources. In contrast, during rare events such as scheduled blackouts \cite{amar:2024}, dimming experiments \cite{barentine:2020} and the decline in business activities during the recent COVID-19 pandemic \cite{li:2022a,bustamante:2021,elvidge:2020a,jechow:2020b}, contributions from different sources of ALAN can be distinguished by careful data examination. Nevertheless, due to their unpredictability, unrepeatability and rarity, the information extracted from these instances is limited. In this context, Earth Hour's coordinated lights-out actions provide a valuable opportunity to fill this gap.

Earth Hour is an annual global campaign organized by the World Wide Fund for Nature (WWF) to raise environmental awareness since 2007. The event involves the voluntary switching off of lights for one hour (20:30-21:30 local time, hereafter) on a selected Saturday in late March. Earth Hour has recently attracted the participation of more than 137 countries and territories~\cite{earthhour:2024}.

Despite the event's popularity and the fact that darkening measurements during lights-out events provide a direct way to quantify the impact of ALAN, surprisingly few researchers have explored this connection. This may be in part due to the potential influence of scattered sunlight, especially in high-altitude locations where twilight can last until after 20:30 in late March.  
To date, only a handful of studies have investigated the effects of Earth Hour, but their results were mixed and observation periods had been short.
For example, a 2018 study in Berlin found a 2-8\% decrease in zenith illuminance~\cite{andreas:2019}. Studies in Spain using Sky Quality Meters (SQMs) had inconclusive results due to sunlight and clouds~\cite{devesa:2016,marco:2012}, and a 2018 report from Yogyakarta, Indonesia was heavily influenced by clouds~\cite{sukma:2019}.

Here, we present the first comprehensive assessment of light reductions during the lights-out event in Hong Kong. This paper will not comment on the effectiveness of Earth Hour in addressing environmental issues. Rather, we combine photographic, video, photometric and spectroscopic observations from the Tsim Sha Tsui (TST) district in Hong Kong, to observe the lights-out and to quantify the changes in night sky brightness and spectral power distributions before, during and after Earth Hour from 2011 to 2024.

\begin{figure} 
	\centering
	\includegraphics[width=\textwidth]{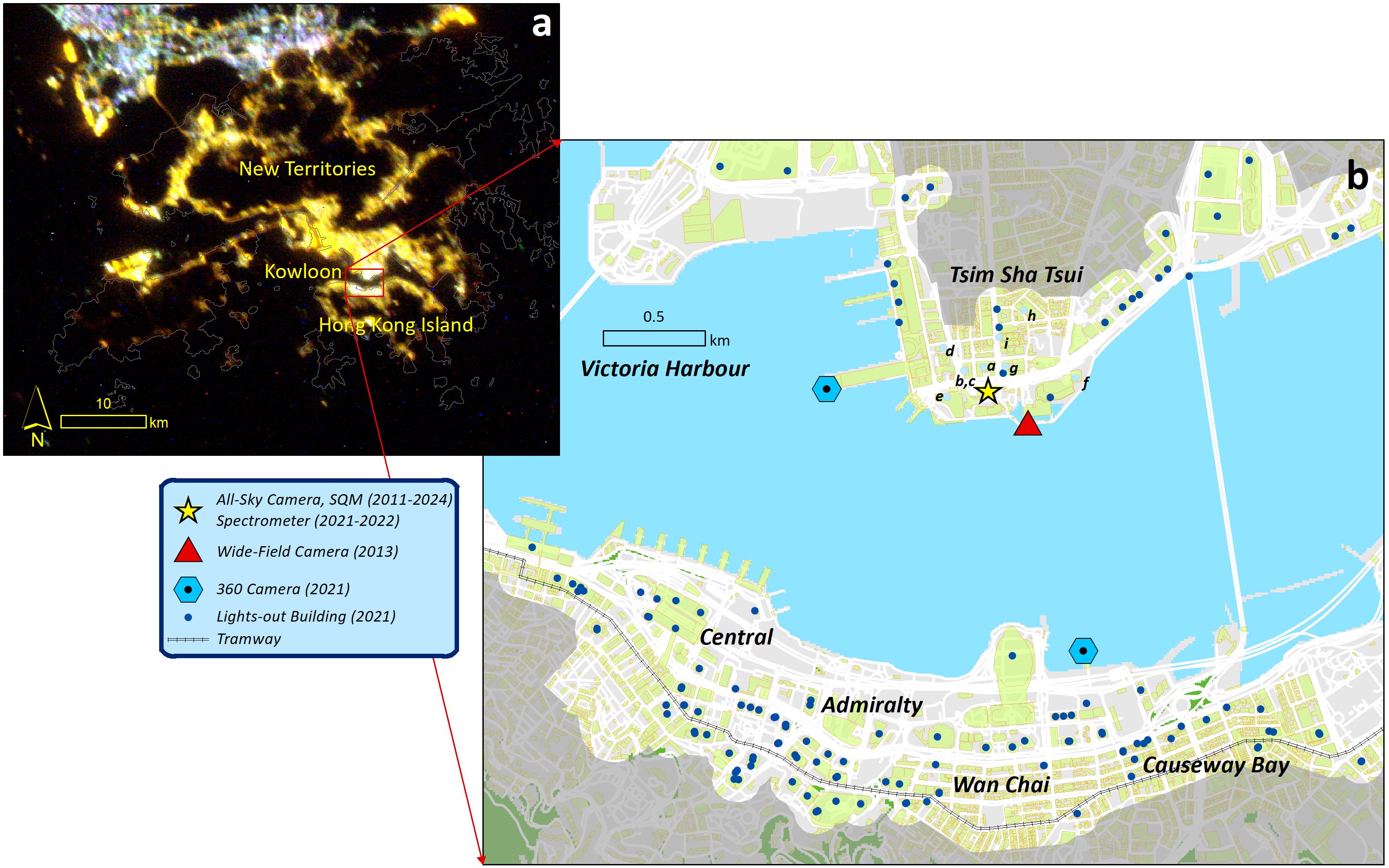} 
	\caption{   
	(\textbf{a}) The nighttime image of Hong Kong taken from the International Space Station at 00:58, 20 January 2015. 
        (\textbf{b}) The 2021 map of downtown Hong Kong, highlighting five CBDs in Kowloon and northern Hong Kong Island. 
        Different symbols represent the locations of observation methods used in this study: 
        a yellow star indicates the same position of the all-sky camera, SQM and spectrometer (at the Hong Kong Space Museum that participates in Earth Hour annually), 
        a red triangle marks the wide-field camera (at the "Avenue of Stars") and
        blue hexagons indicate the positions of the 360-degree cameras. 
        Blue dots represent the buildings (green polygons) that were involved in Earth Hour 2021 and surveyed for this project. 
        Light grey areas indicate unsurveyed regions in 2021.
        The regions surveyed with webcams outside the vicinity of Victoria Harbour are not shown here. 
        The locations of lighting fixtures involved in Earth Hour from 2015 to 2024, as seen in the all-sky images (see Figures~\ref{fig:all_sky_imagesa} to~\ref{fig:all_sky_labels}), are labeled with letters.
        Some of the buildings in northern Hong Kong Island were also involved in Earth Hour 2013 (see Figure~\ref{fig:wide_field_images}).
        The map was created with \textsc{ArcGIS} 10.8 (http://www.esri.com/software/arcgis) using the road network data from \textit{OpenStreetMap}.
        }
	\label{fig:HKmap}
\end{figure}

We select Hong Kong because it has been a leading participant in Earth Hour, with the number of supporting companies, organizations and buildings increased from 3,200 in 2011 to 4,500 in 2020, before declining slightly to 4,000 in 2021-24~\cite{earth_hour_2024,earth_hour_2023,earth_hour_2022}. We choose TST district (Figure~\ref{fig:HKmap}) as the main area to study the impacts of Earth Hour because it is a major commercial and tourist hub in Hong Kong, dominated by decorative and advertising light sources known to contribute significantly to light pollution in the city~\cite{tong:2022,so:2014,pun:2014,pun:2012}.

By applying a "before-and-after" comparison strategy to our multimodal observations, we found that the lights-out of around 120 external lighting of commercial and office buildings in the core business districts (CBDs) would lead to the declines in night sky brightness ranging from 23.4\% to 52.7\% (which are the ratios of the average sky brightness recorded during the lights-out period compared to the averages from 30 minutes before and after the event), illustrating a marked reduction in the intensity of man-made light emissions during the lights-out period. 
These findings provide important insights into the sources and impacts of ALAN, and offer new perspectives on the necessity and benefit of using sustainable lighting practices in urban environments. 

\section*{Results}

\subsection*{Lights-out as seen from all-sky images} \label{sec:all_sky}
\begin{figure} 
	\centering
	\includegraphics[width=0.75\textwidth]{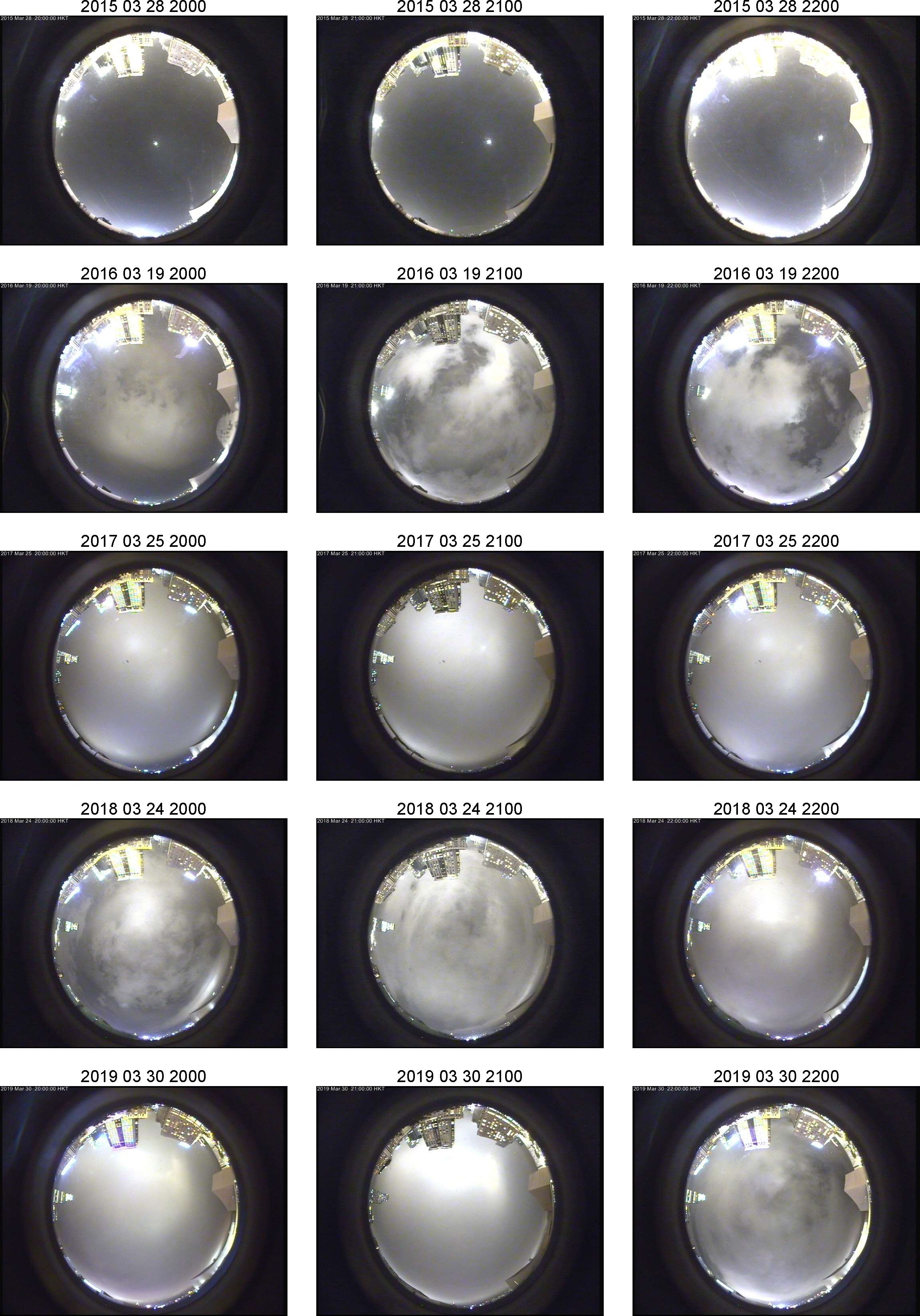}
	\caption{    
	All-sky images of the lights-out in 2015-2019 are arranged from top to bottom. 
        Images were taken at 20:00 (before the lights-out, left column), 21:00 (during the lights-out, middle column) and 22:00 (after the lights-out, right column).
        Timestamps in YYYY MM DD HHMM are repeated above the images. 
        The Moon, if presented and was not blocked by clouds, appeared as a bright spot in the images.
        North is oriented at the top while east is at the left of the images. 
        The all-sky images are courtesy of the Hong Kong Space Museum.}
	\label{fig:all_sky_imagesa}
\end{figure}

\begin{figure} 
	\centering
	\includegraphics[width=0.81\textwidth]{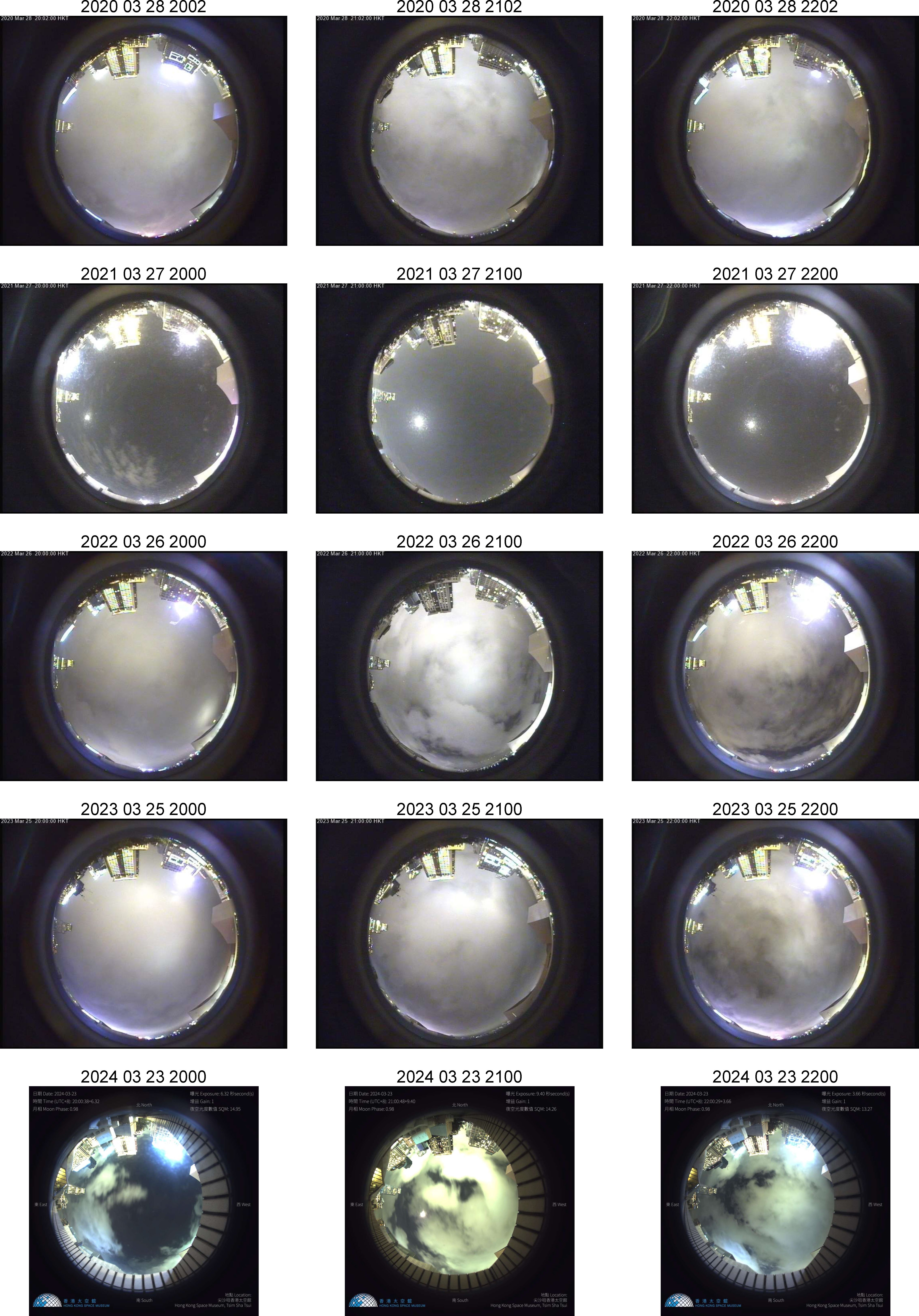}
	\caption{
	Same as Figure~\ref{fig:all_sky_imagesa}, but for 2020-2024. 
        There was a +2 minute discrepancy in the computer clock when these images were saved in 2020. 
        In 2024, a new camera was used, resulting images had higher resolution and some portions of the western and southern horizons were blocked by the roof terrace's sidewall.}
	\label{fig:all_sky_imagesb}
\end{figure}

\begin{figure}
	\centering
	\includegraphics[width=\textwidth]{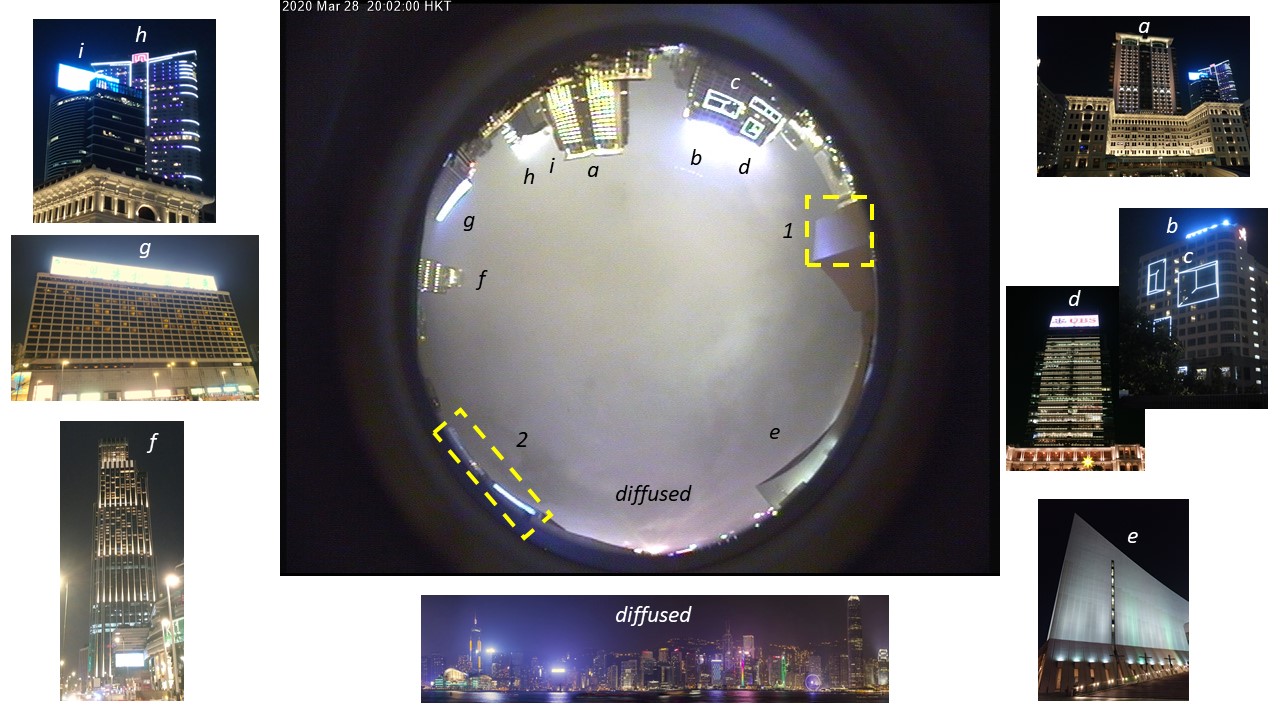} 
	\caption{
        This figure shows a sample all-sky image taken at 20:02 on 28 March 2020 (a half hour before the start of lights-out), with north oriented up and east on the left. Labels are provided clockwise from north, identifying groups of light sources being studied. The side photographs are building images captured in November-December 2020. Figure~\ref{fig:HKmap} shows their locations on the map. The illuminations inside the yellow boxes are reflections (Supplementary Note 2). The all-sky image is courtesy of the Hong Kong Space Museum.}
	\label{fig:all_sky_labels} 
\end{figure}

We examined all-sky camera images captured from the roof of the Hong Kong Space Museum in TST (see Figure~\ref{fig:HKmap} for its location). From the available imagery taken between 2015 and 2024, we selected representative frames at 20:00 (before lights-out), 21:00 (during lights-out) and 22:00 (after lights-out) to qualitatively assess participation in Earth Hour (Figures~\ref{fig:all_sky_imagesa} and~\ref{fig:all_sky_imagesb}). The all-sky images clearly show the reduction in illuminated areas during the lights-out period, with several major commercial and institutional buildings identified as participants (Figure~\ref{fig:all_sky_labels}). Figure~\ref{fig:HKmap} shows their locations on the map.

Our analysis of the all-sky camera images revealed two primary categories of light sources that were impacted during Earth Hour: illuminated building facades and large outdoor billboards. The illuminated facades, labeled $a$, $c$, $e$, $f$ and $h$ in Figure~\ref{fig:all_sky_labels}, corresponded to hotels ($a$ and $c$), a performance facility ($e$) and skyscrapers ($f$ and $h$) in the TST district. These building-mounted lights were typically used for architectural and aesthetic lighting. 
In addition, we identified several large billboards, labeled $b$, $d$ and $i$ in Figure~\ref{fig:all_sky_labels}, which were installed on the roofs of a hotel ($b$) and business towers ($d$ and $i$). 
These billboards, which often adopted the Light-Emitting Diode (LED) technology, were designed to be highly visible from the surrounding harbour area, even during daytime hours. 
In some cases, the billboards appeared merged in the all-sky images due to their proximity (Supplementary Note 1). 
In particular, $g$ refers to a large outdoor LED video wall located on the roof of a hotel, displaying dynamic content. 
The brightness of these kinds of roof billboards and video walls was so intense that they cast strong reflections on nearby structures (Supplementary Note 2). The all-sky camera was unable to capture street-level billboards, such as those on storefronts and illuminated signs near ground level.
We will address the shortcoming with wide-field photographing and a street-level survey in the next sections. 
We also found heterogeneous levels of commitment and participation in the Earth Hour across different light sources and buildings within the TST district over the study period. See Supplementary Note 3 for details.

A prominent light dome was also observed in the southern direction of the all-sky images throughout all years. This light dome is the direct consequence of the scattering of various light sources along the northern Hong Kong Island over Victoria Harbour, at least 1.8~km away. That direction belongs to CBDs such as Central, Admiralty, Wan Chai and Causeway Bay (see Figure~\ref{fig:HKmap} for the map), where many facade lights and billboards are installed and face north. In particular, this light dome dimmed down during the lights-out period in all observed years, suggesting that Earth Hour participation extended beyond the immediate TST district and included the CBDs across the harbour, even though the all-sky camera was unable to capture the individual building-level participation in that area directly.
Again, we address the shortcomings in the following sections. 

\subsection*{Lights-out as seen from wide-field images} \label{sec:2013_dslr} 
During the Earth Hour event on 23 March 2013, a digital single-lens reflex (DSLR) camera with a wide-angle lens was set up at the "Avenue of Stars", a pedestrian walkway located along the Victoria Harbour waterfront in TST, approximately 600 m northeast of the all-sky camera location (Figure~\ref{fig:HKmap}). The camera was oriented towards the south-southwest, capturing a horizontal field of view of approximately $110^{\circ}$, which allowed it to document the lights-out and lights-on events of the external lighting in northern Hong Kong Island every 20 seconds from 17:02 to 22:31.

\begin{figure}
	\centering
	\includegraphics[width=\textwidth]{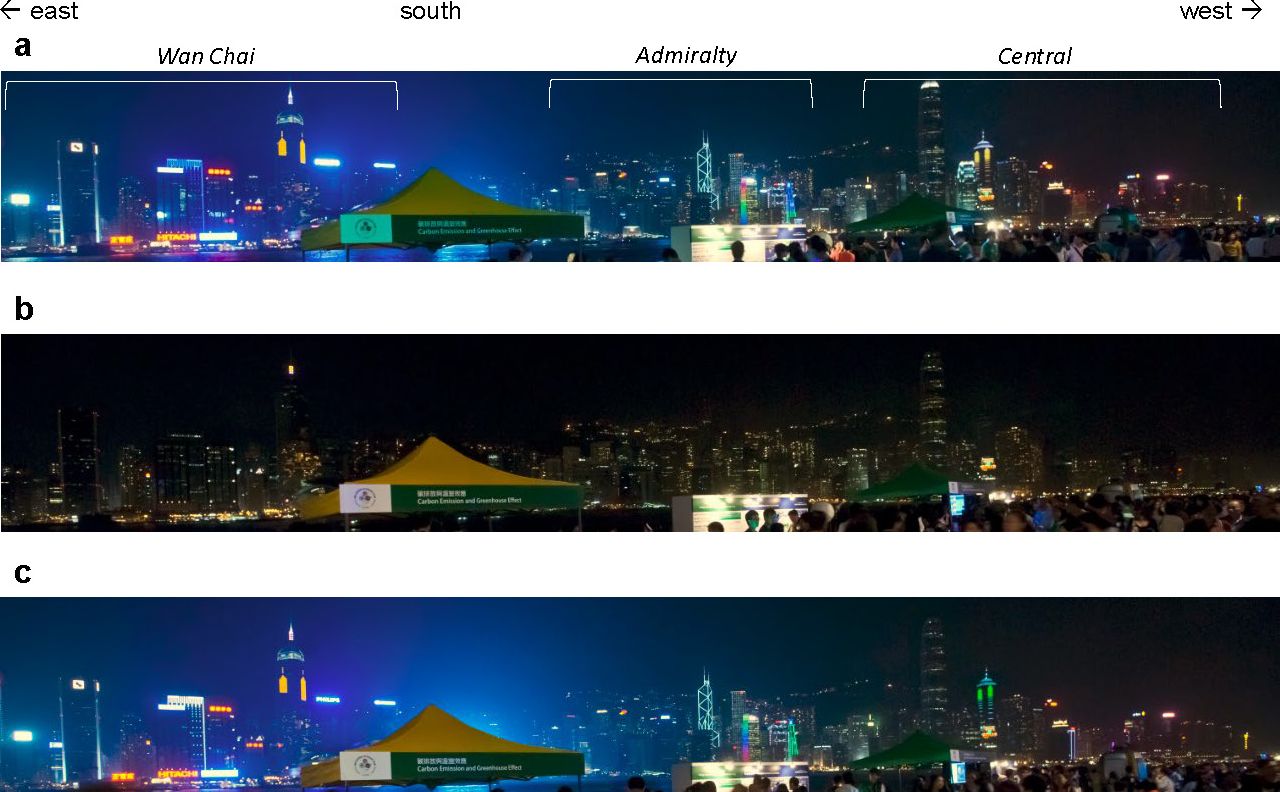} 
	\caption{
        Cropped wide-field images of CBDs taken at TST during Earth Hour 2013: 
        (\textbf{a}) 20:00 (before the lights-out). 
        (\textbf{b}) 21:00 (during the lights-out). 
        (\textbf{c}) 22:00 (after the lights-out). 
        A booth canopy blocked a part of Wan Chai in the foreground.
        See Figure~\ref{fig:HKmap} for the locations of the districts labeled.}
	\label{fig:wide_field_images} 
\end{figure}

The wide-field images revealed that the external lighting that turned off during the Earth Hour event was predominantly composed of illuminated billboards and facade lights on skyscrapers and business towers in the CBDs along the northern shore of Hong Kong Island, including Wan Chai, Admiralty and Central (see Figure~\ref{fig:HKmap} for their locations). The "before-and-after" comparisons of images taken at 20:00 (before the lights-out), 21:00 (during the lights-out) and 22:00 (after the lights-out), as shown in Figure~\ref{fig:wide_field_images}, provide visual evidence to support this observation.

\subsection*{Lights-out as identified from a comprehensive survey} \label{sec:2021_survey} 
To expand our observations beyond the direct views from both sides of Victoria Harbour, highlight the impact of decorative and advertising lighting, and assess the overall nighttime activities, we conducted a comprehensive event participation survey in the CBDs during Earth Hour 2021. Aiming to identify individual external lighting that turned off during the lights-out period as comprehensively as possible, the survey was conducted by taking videos from the tramway, stationing and photographing at different locations with smartphones, DSLR and 360-degree cameras. We also examined the footage provided by WWF-Hong Kong, as well as webcam images for road traffic and weather monitoring distributed across the entire city by the government.
See Ref. \cite{liu:2023a} for details.

Some survey results are presented in Figure~\ref{fig:HKmap}. In sum, we identified 122 individual lighting fixtures that participated in the lights-out. The participation was quite complete among commercial buildings in the CBDs, especially those with facade illuminations and billboards facing Victoria Harbour. On the other hand, we found that some owners of billboards facing the inner streets near street level and most of the buildings outside the CBDs did not support the event. These findings suggest that while commercial entities in Hong Kong's leading business districts demonstrated a high level of commitment to the Earth Hour initiative, there is still room for improvement in engaging the owners of billboards and buildings outside of the CBDs to further expand the impact of the event.

\subsection*{Sky darkening as seen from NSB observations} \label{sec:nsb_results} 
\begin{figure}
	\centering
	\includegraphics[width=0.5\textwidth]{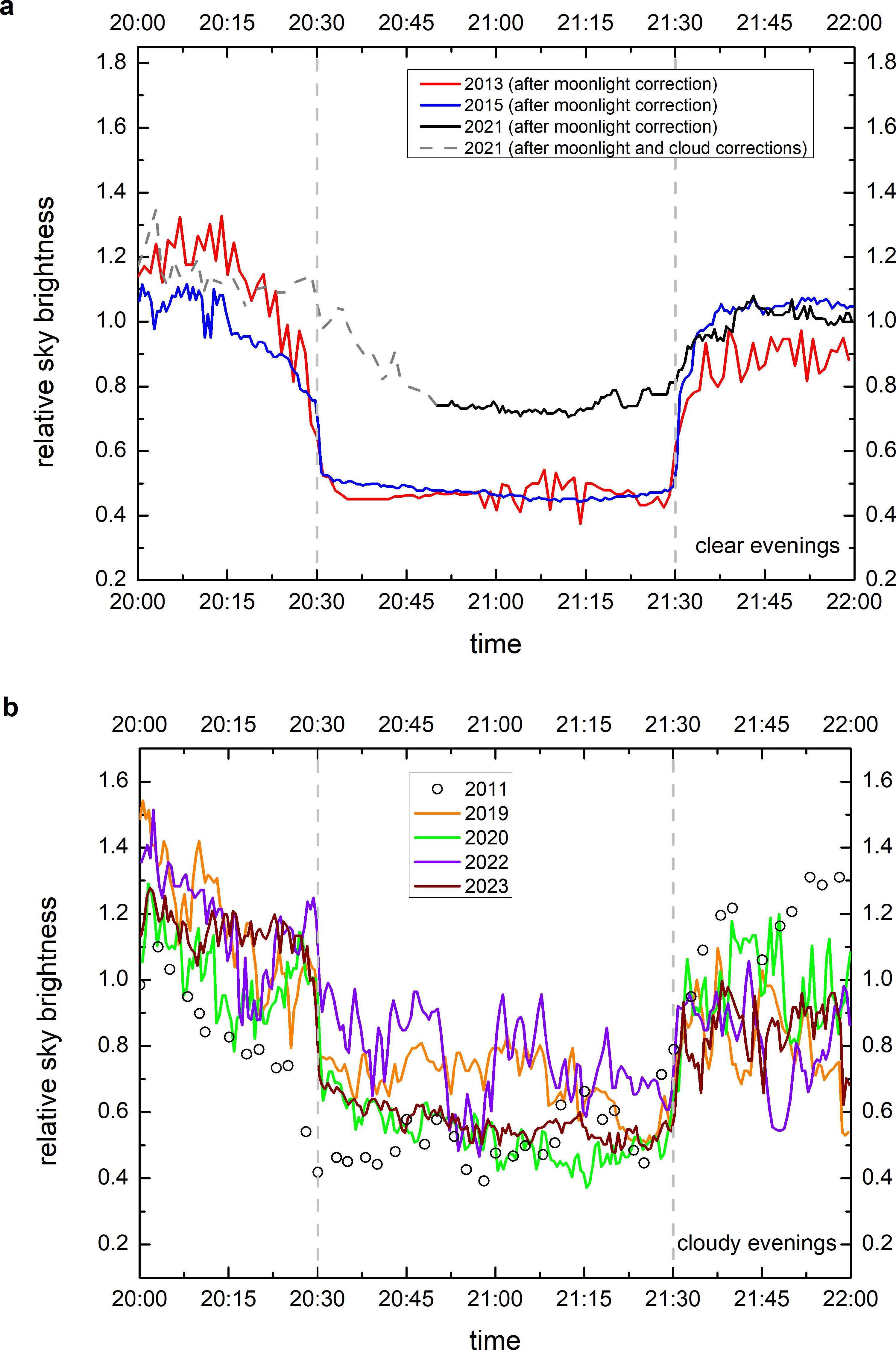} 
	\caption{
        (\textbf{a}) Relative sky brightness in term of NSB at TST in Hong Kong during clear Earth Hour evenings. 
        (\textbf{b}) Same for cloudy evenings. 
        Each NSB measurement was converted to a linear quantity with Equation~\ref{eq:log2linear} in Methods then presented as a ratio with reference to the average brightness measured 30 minutes before and after the event each year.
        The 2011 light curve is plotted as circles to indicate its sparse temporal sampling resolution (150 seconds per data point) compared to the others (60 or 30 seconds per data point). 
        The 2013, 2015 and 2021 light curves have been corrected for moonlight (see Methods for details).
        The 2011, 2013 and 2015 light curves have also been corrected for shielding on the sensor's housing (Supplementary Note 9 and Supplementary figure S6). 
        The 2021 data collected before 20:50 and corrected for cloud (Supplementary Note 8, Supplementary figures S3-S5) and moon contributions are shown here in a gray dashed curve for completeness only.
        Vertical dashed lines mark the start and end of the lights-out period.}
	\label{fig:TST_nsb} 
\end{figure}

Building on our understanding of ALAN lights-out participation from previous sections, we quantify the influence of the event on the night sky in terms of its brightness variations in this section.
Night sky brightness (NSB) has been measured over the roof of the Hong Kong Space Museum (co-located with the all-sky camera) in TST since 2011. All of our NSB measurements were conducted without sunlight contamination, as evening astronomical twilight ended at least 30 minutes before the lights-out event.
We present the NSB light curves observed during Earth Hour 2011, 2013, 2015, 2019, 2020, 2021, 2022 and 2023 in Figure~\ref{fig:TST_nsb}. These data are segregated by clear and cloudy observational conditions. Earth Hour NSB data are unavailable for 2014 and 2017 due to sensor outages. Additionally, the data collected during Earth Hour 2012, 2016, 2018 and 2024 were found to be excessively noisy owing to substantial cloud cover fluctuations, preventing the detection of lights-out impacts (Supplementary Note 4).

As revealed by the all-sky image sequence in Supplementary figure S2, the 2021 observations prior to 20:50 were influenced by clouds. This influence was more significant as the cloud patches approached the zenith at TST between 20:31 and 20:43. We have removed the cloud influence from the NSB observed before 20:50 by linearly interpolating the cloud-free NSB readings during the cloudy periods (Supplementary Note 8, Supplementary figures S3-S5). In Figure~\ref{fig:TST_nsb}, we have included the moonlight- and cloud-corrected NSB data before 20:50 for visualization purposes. Furthermore, the 2013, 2015 and 2021 observations have been corrected for the influence of moonlight (see Methods for details).

As seen in Figure~\ref{fig:TST_nsb}, a clear drop in NSB is evident during the lights-out period on each light curve. Specifically, the NSB dropped when the lights were turned off at 20:30 and then rose again when the lights were turned back on at 21:30. These timings of the NSB changes align with the observations from the all-sky and wide-field images mentioned above. 

Quantitatively, by comparing the average sky brightness during the lights-out period with the averages from 30 minutes before and after the event, the zenith sky was darkened by 52.7\%, 52.2\% and 25.5\% during the clear sky periods in 2013, 2015 and 2021, respectively. For 2021, we only used the moonlight-corrected NSB recorded from 20:50 onward in these calculations. The lower degree of darkening observed in 2021, amidst the COVID-19 pandemic, appears to be attributable to the reduction in ALAN to save on electricity costs or due to the closure of businesses, as happened in other locations reported in the literature \cite{li:2022a,bustamante:2021,elvidge:2020a,jechow:2020b}.

The zenith sky was darkened by 47.8\%, 30.3\%, 46.5\%, 23.4\% and 42.9\% under cloudy conditions in 2011, 2019, 2020, 2022 and 2023, respectively. Research has shown that clouds can dramatically enhance the back-scattering of upwelling ALAN~\cite{arroyo:2024,sciezor:2020b,so:2014,kyba:2012,christopher:2011}. In general, at a location affected by light pollution, a cloudy sky will appear brighter compared to clear sky conditions, and vice versa. Specifically, a previous study in Hong Kong found that urban NSB can be brightened sixfold as cloud coverage increases from 40\% to 85\%~\cite{so:2014}. For the cases where it was cloudy during the lights-out period (i.e., 2011, 2019, 2020, 2021 before 20:50, 2022 and 2023), the sky would have been brightened due to the (uneven) back-scattering of ALAN from the clouds, partially masking some of the Earth Hour's impacts. On the other hand, we were able to depict the Earth Hour's impacts more accurately under clear sky conditions in 2013 and 2015. Indeed, the sizes of the darkening (around 50\%) in these clear years were comparable, and the sky brightness generally "returned" to the pre-Earth-Hour level when the lighting resumed.

We are uncertain how the COVID-19 pandemic may have influenced the sky in 2020 and 2022, as these years were cloudy and the real impact, if any, could have been hidden behind the cloud fluctuations. A separate study that analyzes a broader dataset beyond Earth Hour data would be better suited to investigate the full impacts of COVID-19 in different years.

\subsection*{Man-made emissions darkening as seen from spectroscopic observations} \label{sec:spectral_results} 
\begin{figure}
	\centering
	\includegraphics[width=0.65\textwidth]{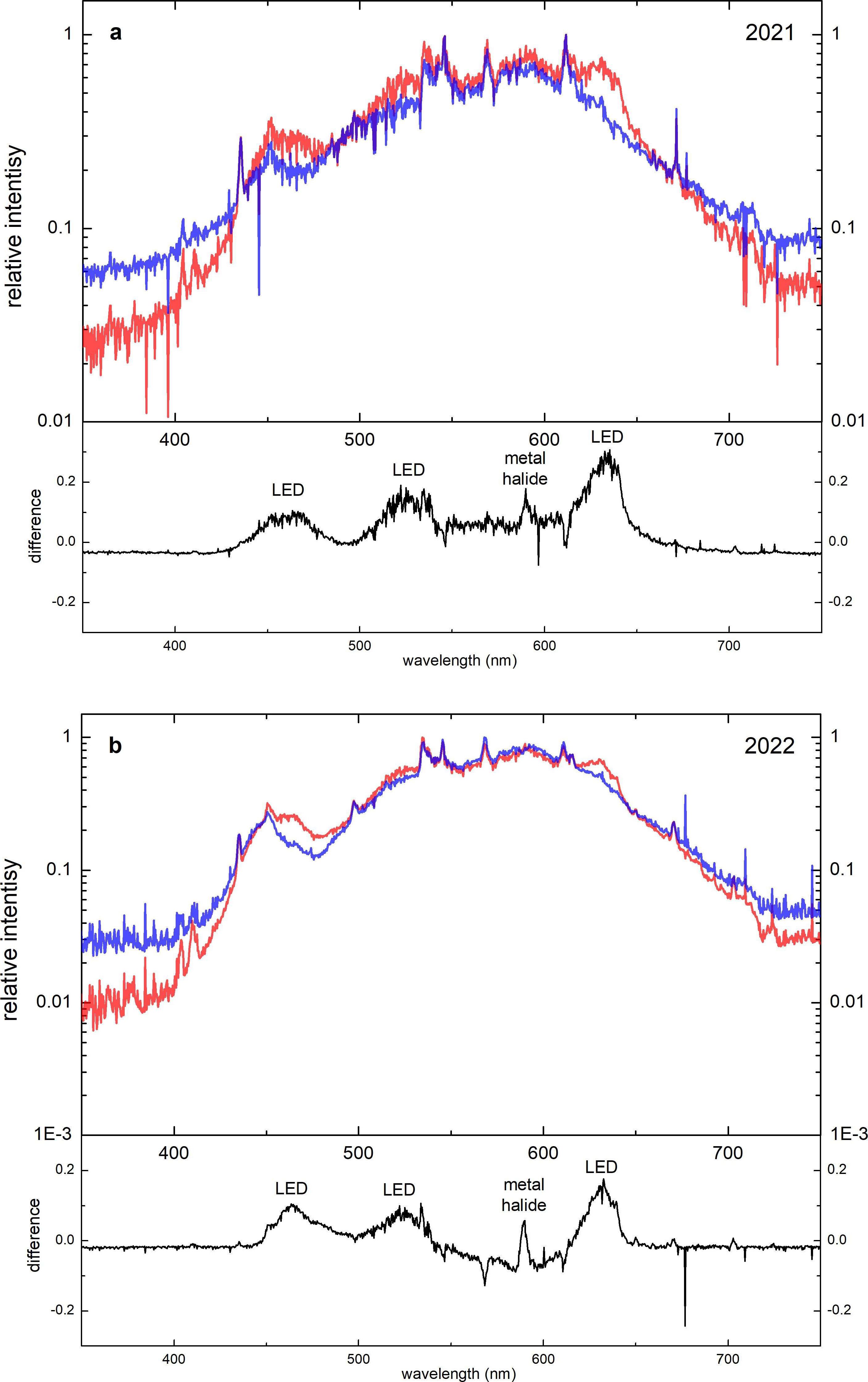} 
	\caption{
        (\textbf{a}) Zenith sky spectra, plotted in logarithmic scales, acquired at TST (same location as the all-sky and NSB observations) in Hong Kong during the 2021 Earth Hour evenings.
        (\textbf{b}) Same for 2022. 
        The red curves represent the median "before lights-out" normalized spectra, while the blue curves show the "after lights-out" version. The black curves, plotted in linear scales, illustrate the "before-minus-after" difference, with the major peaks labeled according to the types of light sources.}
	\label{fig:spectra} 
\end{figure}

Our NSB observations detected changes in sky brightness irrespective of the light sources. To identify which types of lighting contributed to the sky darkening, we set up a portable spectrometer in TST (same location as the all-sky and NSB observations) to record the sky spectra during the Earth Hour evenings of 2021 and 2022. From the acquired data, we compared the spectra taken during the 20:00-20:25 period with those taken during the 20:45-21:15 period, as a "before-and-after" residual analysis (Figure~\ref{fig:spectra}).

If lights-out had not occurred and lighting and atmospheric conditions had remained largely unchanged during the period, the residuals would have been negligible. However, we detected noticeable residuals attributed to the lights-out. These residuals are particularly pronounced in the ranges of 445-500, 500-540 and 615-650~nm, corresponding to the dimming of blue, green and red emissions from LED video walls (such as $g$ seen in Figure~\ref{fig:all_sky_labels}). 
Additionally, in the range of 585-595~nm, we observe a marked narrow emission feature from metal halide floodlights used for facades and billboards. 
In contrast, emissions from primarily the high-pressure sodium lamp poles that remained on throughout the evenings, are not present in the residuals.

The 2022 residuals are smaller near the 550-610~nm region than those of 2021. By checking the all-sky images (Figure~\ref{fig:all_sky_imagesb}) and confirming through follow-up on-site spectroscopic measurements, it was found that the facade lighting $e$ of the performance facility did not switch off during Earth Hour 2022, resulting in a smaller "before-and-after" residual near that particular spectral region. This suggests that the changes in the sky spectrum during the lights-out period were sensitive to the specific lighting sources that were dimmed or turned off in the local area. The persistence of certain lighting fixtures, like the facade lighting in 2022, led to a less pronounced change in the overall sky brightness and spectral composition compared to the other years.

Our observations support the conclusion that the sky darkening we observed during the Earth Hour evenings in the previous sections is directly attributed to the switching off of decorative and advertising lighting fixtures in the local area (at least for 2021 and 2022). The persistence or absence of certain spectral signatures in the "before-and-after" comparisons provides strong evidence that the lights-out event was the primary driver of the observed changes in the night sky brightness and spectral composition.

\section*{Discussion} \label{sec:discussion}
Rather than waiting for the next blackout to occur at an unpredictable and infrequent time and place, Earth Hour sets the stage for us to directly investigate the impacts of ALAN on the environment each year.
This repeatable nature of the event not only allows us to validate previous observations but also facilitates the discovery of new trends and phenomena. Such repeatability is especially crucial given the varying weather and lunar conditions that can influence the measurements.
While this paper does not aim to evaluate the overall success of Earth Hour, we focus on verifying and quantifying the impact of ALAN on the environment through various observational methods, utilizing data from the past 14 years of events.

The lights-out causes a reduction of up to 50\% in the brightness of the zenith sky. 
This level of reduction is significantly higher than that observed in other cities, such as Berlin, where the reductions range from 2\% to 8\% \cite{andreas:2019}. Apart from the degree of lights-out engagement, the discrepancy in these findings can be attributed to the difference in building density and lighting intensity between locations (compare Figure~\ref{fig:HKmap} in this work and their Figure 1). In cities like Berlin, where building lighting is likely less dense and less intense, the overall contribution to sky brightness is reduced, leading to a smaller impact during similar events. In contrast, the substantial darkening observed in Hong Kong can be largely attributed to the simple act of turning off approximately 120 decorative or advertising lights on commercial buildings and shopping malls in the vicinity of the observation location. 
Although they represent only a small fraction of the total external lighting in the city and are mostly concentrated within the CBDs, our findings indicate that they are in fact the main contributors to light pollution in Hong Kong.
Furthermore, the transient nature of the darkening\textemdash where sky returns to the pre-event brightness level after the lights are turned back on\textemdash emphasizes the need for more permanent solutions. 

Our findings highlight clear opportunities for potential sustainable lighting solutions. 
The significant sky darkening during Earth Hour suggests that limiting decorative or advertising lighting, especially in commercial districts, can have an immediate and significant impact. Policies that mandate turning off these lights after certain hours, similar to the current \textit{Charter on External Lighting}~\cite{law:2024}, could be expanded to ensure more consistent reductions. Limiting the installation of decorative and advertising lighting in key areas could also reduce the impact of artificial lighting on residential zones and natural habitats.
In addition, our spectral analysis shows that high-intensity light sources, especially LED building facades, billboards and floodlights, contribute heavily to light pollution. Retrofitting them with phosphor-converted amber (PC amber) LEDs, which emit less blue light, would significantly reduce the skyglow while maintaining energy efficiency~\cite{aube:2016led}. Light sources with the right choice of spectral characteristics are equally important from physiological and ecological perspectives~\cite{wu:2023,mcnaughton:2021b,aube:2013}. 
Other achievable means include reducing the intensity of the lamp, shielding the cobrahead by cutoff design to avoid light spill, installing motion detectors and planting a row of trees next to the walkway~\cite{aube:2016led,aube:2015}. According to a model simulating the propagation of ALAN in the nocturnal environment, the adverse potential impacts would be reduced by up to nearly 2,000 times if all actions were taken~\cite{aube:2016led}. Ideally, concerning the positive sustainability outcomes, such effective strategies should be included in national environmental protection and territorial development policies. 

Our study has two main limitations. First, our sky brightness estimations may not perfectly capture the impact of ALAN as perceived by humans, for two reasons: 1) there is a difference in spectral sensitivity between the photometer (SQM) used in our measurements and the human eye (i.e., scotopic vision)~\cite{miguel:2017}; and 2) the SQM pointed towards the zenith and measured the intensity of the scattered light from the overhead sky, which was illuminated by upward-facing light sources and upward-reflecting light. However, some external lighting, such as video walls and billboards, is intentionally designed to illuminate horizontally to attract people's attention or be visible over longer distances. As a result, human observers may experience brighter illumination before the lights are extinguished, and therefore a more pronounced darkening during the lights-out period, compared to our sky brightness measurements.

Second, the ongoing transition to LED illumination is altering the spectral composition of artificial skyglow, in particular by increasing the proportion of blue light emission~\cite{arroyo:2024,levin:2023a,robles:2021,kyba:2017}. Depending on factors such as the application of LED illumination (e.g., total amount, intensity and upward light output), measurement distance (as blue light scatters more over longer distances) and atmospheric optics (e.g. cloud cover), the SQM may record a brighter or darker NSB reading, as the data interacts with the changing sky spectrum in a complex manner~\cite{miguel:2017}. Although our spectroscopic observations have clearly identified the contribution of LEDs to the brightening of the night sky, future analyses should carefully consider the effects of the ongoing transition in lighting technology and the resulting changes in sky color when interpreting NSB measurements~\cite{arroyo:2024}.

Our study utilized multiple data sources, including all-sky images, wide-field images, and both photometric and spectroscopic data. Future work includes focusing on cross-verification and cross-calibration among these different sources to enhance the validity of our findings.

Remote sensing of night lights from space has become a popular way to study light pollution and has greatly improved our understanding of the impacts of ALAN on the environment from regional to global perspectives\cite{levin:2020}. Compared with ground-based measurements like our study, remote sensing measurements are superior in geographic coverage (i.e., a single data frame covers a large landmass) but are limited in temporal extent (e.g. one of the popular choices, Suomi National Polar-Orbiting Partnership satellite, which equips the Visible Infrared Imaging Radiometer Suite (VIIRS) instrument for nighttime imaging, overpasses a spot after local mid-night once only every night~\cite{liu:2025,liuhku:2021,miller:2012}; nighttime DSLR images captured by astronauts aboard the International Space Station also facilitate ALAN studies~\cite{miguel:2019a}, although they were captured at irregular intervals). Infrequent satellite visits and occasional photographing pose a challenge in detecting short-term light variations such as the hourly ALAN changes during Earth Hour and unexpected blackouts~\cite{levin:2025}. 
In addition, remote sensing detects the scattered and upward ALAN from the ground to space, often at a sparse resolution that makes individual light sources indistinguishable. This contrasts with the direct impacts of ALAN from specific external lighting as recorded by ground-based sensors, exemplified in the present study. 
Although the popularity of space-based remote sensing of night lights is growing and improvements have been made (e.g., higher spatial resolution satellite data are available nowadays~\cite{liuhku:2024,liu:2023b,li:2018,zheng:2018}) or called for~\cite{kyba:2024,elvidge:2007}, ground-based measurements continue to play an unparalleled role in examining horizontal emission that is of particular concern to people, such as light from illuminated billboards shining into bedrooms~\cite{levin:2025}. 

During the lights-out in 2021 when the darkest sky was recorded at TST, after removing the contribution from moonlight, the sky near the zenith was still around 50 times brighter than the natural night sky brightness in the Johnson \textit{V}-band~\cite{masana:2021}. While this comparison is an approximation due to spectral mismatch (Supplementary Note 11), it offers an order-of-magnitude estimate of how a city's light pollution could be improved when decorative and advertising lights are turned off simultaneously. Even at this reduced level of ALAN, the Milky Way would still be invisible to the naked eye at TST, as the dark adaptation of the eye could not be fully achieved~\cite{so:2014}.
Nevertheless, the evident darkening in the case of Hong Kong, has surprised many by demonstrating the continued allure of nocturnal cityscapes even without anthropogenic illumination. 
If the lights-out were more thoroughgoing, similar to the blackout experienced in some cities \cite{joshua:2025,jane:2010}, or if sustainable lighting measures were in place, the potential to observe and appreciate the natural night sky could be significantly enhanced. Such conditions would offer an opportunity for urban dwellers to reconnect with the beauty of the cosmos, fostering a greater appreciation for both the environment and the importance of reducing light pollution.

\section*{Methods}

\subsection*{Observing location} \label{sec:observing_location}
All sky conditions and night sky brightness (NSB) were monitored on the roof of the Hong Kong Space Museum (22.29$^{\circ}$ N, 114.17$^{\circ}$ E) located in Tsim Sha Tsui (TST), Kowloon, Hong Kong. TST represents one of the densely populated downtown districts of Hong Kong, characterized by a proliferation of high-rise hotels and commercial buildings. This urban area features extensive external lighting, including large illuminated billboards on roofs and walls, as well as LED displays with animated effects, operating at high intensities to attract attention from across the harbour and stand out from surrounding illuminators. Additionally, many buildings in TST employ strong facade lighting to highlight architectural details. Wide-field imagery was captured at the popular tourist attraction "Avenue of Stars", a promenade along the Victoria Harbour waterfront in TST, situated approximately 600 m northeast of the museum. The locations of these observation sites are shown in Figure~\ref{fig:HKmap}.

At the time of writing, Hong Kong has not adopted any mandatory regulations governing the use of external lighting. Instead, the Hong Kong government has implemented a voluntary scheme, the \textit{Charter on External Lighting}, since 2016, which invites owners and responsible parties to switch off decorative, promotional or advertising lighting installations at 22:00 (a stricter requirement introduced in 2023), 23:00 or 00:00 till 07:00 the following morning~\cite{law:2024}. As the observations in this study were made before 22:00, we do not expect the voluntary \textit{Charter} to have influenced the recorded lighting conditions.

\subsection*{All-sky images} \label{sec:all_sky_method}
The Moonglow Technologies all-sky camera was used to capture all-sky images before 2024. This camera employs a 1/3 inch Sony Super HAD CCD II color sensor paired with a 1.24~mm F/2.8 fisheye lens, yielding an effective pixel size of 546$\times$457 in the 190$^\circ$ hemispherical field of view.
All-sky image data has been available since August 2014. Prior to 2020, images were archived on an hourly basis at the top of each hour (e.g. 20:00, 21:00, 22:00, etc.) due to software constraints. Since 2020, the camera has been configured to record images at a one-minute cadence.

In 2024, the museum tested a new self-developed camera system. The camera employs a 16-mm diagonal Sony IMX 533 CMOS color sensor paired with a 2.7~mm F/1.8 fisheye lens, yielding an effective pixel size of 3,008$\times$3,008 in the 185$^\circ$ hemispherical field of view. The camera took images at a one-minute interval. During the testing stage, some of the horizon views in the west and south were blocked by the roof terrace sidewall. 

During operation, both cameras automatically adjusted settings such as exposure time to optimize visibility of sky conditions. These settings details were not recorded in the image metadata, preventing reliable quantitative comparisons between images. Additionally, some images exhibited saturation near external light sources, obscuring information on actual light intensity. 

For the purposes of this study, we obtained the images captured during the evenings of Earth Hour events from the Hong Kong Space Museum. 

\subsection*{Wide-field images} \label{sec:wide_field_method}
Wide-field images were captured using a Sony DSLR-A900 camera during Earth Hour 2013. These images were acquired at an ISO of 1,600, shutter speed of 1/6 seconds and aperture of F/6.3, with the camera mounted on a tripod and the flash disabled. The lens had a focal length of 16~mm. Similarly to the all-sky images, the digital single-lens reflex (DSLR) camera was not radiometrically calibrated, preventing a quantitative analysis of the lighting changes during the Earth Hour event.

\subsection*{A comprehensive survey in 2021}\label{sec:2021_method}
To investigate the impact of Earth Hour 2021 on lighting levels within the core business districts (CBDs) surrounding Victoria Harbour in Hong Kong, we employed a multipronged observational approach. Given the time-limited nature of the event and resource constraints, it was not feasible to make comprehensive direct observations at all street corners within the target areas. Instead, we used a variety of data sources and survey techniques to fully document the spatial extent of lights-out participation.
First, a smartphone camera was mounted on the upper deck of a tram traversing the northern shore of Hong Kong Island from Sheung Wan to Causeway Bay. The journey took about 45 minutes. This mobile video platform allowed us to record the surrounding environment before, during and after the Earth Hour event. Comparative analysis of video footage captured on the Earth Hour evening versus the prior evening facilitated the identification of specific lighting fixtures that were extinguished.
Second, we deployed volunteer observers equipped with DSLR cameras and 360-degree cameras to strategic vantage points, such as hilltops and observation piers in Tsim Sha Tsui and Wan Chai. These stationary survey locations provided wide-angle views of the cityscape.
Third, another group of volunteer observers traversed the inner street networks (away from the tramway) of select CBDs, including Tsim Sha Tsui, Tsim Sha Tsui East, Mong Kok, Central and Causeway Bay. This ground-level mobile survey utilized smartphones, DSLRs, 360-degree cameras and action cameras to document lighting conditions at a granular scale.
We trained the volunteers and maintained close communication with them during the survey to ensure their data was suitable for analysis.
Finally, we obtained documentary footage recorded by WWF-Hong Kong during the Earth Hour event to confirm some of the findings from the in situ surveys.
To investigate the impact outside the CBDs, we also accessed government-operated webcam feeds that monitor road traffic and weather conditions of the entire city to supplement observational data.
Through the integration of these multimodal data sources, we were able to identify 122 individual lighting fixtures that participated in Earth Hour 2021 (Figure~\ref{fig:HKmap}).
See Ref. \cite{liu:2023a} for additional details.

\subsection*{NSB observations} \label{sec:nsb_obs_method}
There are many ways to quantify NSB~\cite{kocifaj:2023d}. In this study, the zenith sunlight-free NSB data presented were obtained through the monitoring efforts of two established networks: the \textit{Hong Kong Night Sky Brightness Monitoring Network} (NSN), operational from 2010 to 2014, and the \textit{Globe at Night - Sky Brightness Monitoring Network} (GaN-MN), which has maintained continuous observations since 2014.
Instrumental measurements of NSB were conducted using Sky Quality Meter - Lens Ethernet (SQM-LE) devices, a class of semiconductor-based photometers. These SQM-LE units recorded NSB at intervals of 30 seconds (for Earth Hour events in 2015-2016, 2018-2024), 60 seconds (2012 and 2013) or 150 seconds (2011). The designation "L" denotes that these sensors incorporate a lens element, which narrows the full width to half the maximum (FWHM) of the angular sensitivity to approximately 20$^{\circ}$. In addition, ethernet connectivity (the "E" designation) of these instruments facilitates automated data transmission.
SQM devices have emerged as a widely adopted tool for quantifying light pollution, with applications documented both locally within Hong Kong~\cite{so:2019,so:2014,pun:2014,pun:2012} and globally~\cite{hanela:2018}. The spectral sensitivity, angular response, photometric accuracy, linearity, temperature stability and other long-term performance characteristics of these sensors have been extensively documented ~\cite{fiorentin:2025,fiorentin:2022b,bara:2021,salvador:2019,miguel:2017,pravettoni:2016,outer:2015,schnitt:2013,peter_outer:2011,sqmreport,sqmlreport}.
We routinely inspect and calibrate the SQM instruments used in this study. As the NSB values reported here represent relative measurements from a given meter, corrections for sensor and glass window aging, as described in previous studies~\cite{fiorentin:2025,bara:2021,so:2014}, were not applied in the present analysis. We have also ignored other factors such as changes in the sensor configurations and atmospheric optical properties over the years because they have minor impacts on our observations (Supplementary Notes 9 and 10).

The SQM instruments utilized in this study report NSB measurements in the astronomical logarithmic unit of magnitude per square arcsecond (mag~arcsec$^{-2}$). In this convention, a brighter night sky corresponds to a smaller numerical value, and vice versa.
When quantifying the zenith sky darkening induced by lights-out participation during Earth Hour, we first converted the logarithmic SQM readings ($M$ in mag~arcsec$^{-2}$) to a linear luminance quantity ($L$ in cd m$^{-2}$) using the expression~\cite{hanela:2018} (Supplementary Note 11):
\begin{equation}
L = 10.8 \times 10^4 \times 10^{-0.4 M}. \label{eq:log2linear}
\end{equation}
For each evening, then we compute the ratio of the average sky brightness recorded during the lights-out period ($\bar{L}_{\text{EH}}$) to the average brightness measured 30 minutes before and after the event ($\bar{L}_{\text{non-EH}}$) as: 
\begin{equation}
 \frac{\bar{L}_{\text{EH}} - \bar{L}_{\text{non-EH}}}{\bar{L}_{\text{non-EH}}} \times 100\%, \label{eq:sqm_darkening}
\end{equation}
which is the zenith sky darkening percentages.
For the 2021 observation campaign, only cloud-free brightness data obtained from 20:50 onward were considered in the analysis.
If we had instead calculated the darkening relative to the average sky brightness recorded solely in the 30-minute period following the lights-out, as was done for the 2021 data, the resultant darkening percentages would be 55.8\%, 45.9\%, 53.3\%, 15.2\%, 45.6\%, 6.3\% and 32.9\% for the years 2011, 2013, 2015, 2019, 2020, 2022 and 2023 respectively.

During the course of Earth Hour events from 2011 to 2024, we collected a total of 115 individual NSB datasets from 21 distinct locations in Hong Kong (NSB data for 2014 and 2017 are unavailable). We treated the lights-out period from 20:30 to 21:30 and the subsequent lights-on period as independent events, partitioning each dataset into two segments: 20:00-21:00 and 21:00-22:00.
After implementing data quality control measures to discard any split datasets containing zero readings (due to instrumental or communication issues), a total of 216 valid datasets remained after the partitioning process. These datasets were collected at a diverse range of sites, ranging from urban centers and residential neighborhoods to dark country parks and specialized land uses (e.g. wetlands, container terminals, airport)~\cite{so:2014,pun:2014}.
We then isolated the valid datasets exhibiting potential changes in NSB in response to the 20:30 lights-out or 21:30 lights-on events (classified as positive cases for further analysis) from those that did not demonstrate such changes (negative cases). This categorization was performed using the methodologies detailed in Supplementary Notes 5 to 7 and Supplementary figure S1.
With the exception of TST, all other locations do not observe significant reductions in sky brightness.
The present analysis focuses on the positive case datasets obtained from TST during the Earth Hour events of 2011, 2013, 2015, 2019, 2020, 2021, 2022 and 2023.

\subsection*{Removing moonlight from NSB observed} \label{sec:nsb_moon_method}
The Moon was present above the horizon during Earth Hour events in 2013, 2015, 2020, 2021 and 2023 (see Supplementary table S5 for details on the prevailing Moon conditions). If the Moon was not entirely obscured by cloud cover, the direct and scattered moonlight falling within the field-of-view of the employed SQM sensors may have influenced the recorded NSB measurements.
Based on an analysis of all-sky and wide-field imagery, sky conditions were classified as clear in 2015, partially cloudy in 2013 and 2021, and overcast in 2020 and 2023. Therefore, we assume that moonlight impacted the NSB observations in 2013, 2015 and 2021 only.

To estimate the contributions of moonlight to the observed NSB, we utilized a moonlight model~\cite{so:2014,krisciunas:1991} that calculates the brightness of direct and scattered moonlight ($I_{\text{moon}}$) as a function of the lunar phase angle ($\alpha$, accurate to 0.2$^\circ$) and the angular separation ($\phi$) between the observed sky patch and the Moon. As the SQM sensors were pointed at the zenith, the value of $\phi$ was set equal to the Moon's zenith angle ($\phi_{\text{moon}}$, accurate to 1$^\circ$). Lunar properties were calculated using \textsc{Alcyone Ephemeris} version 4.3. 
The original moonlight model~\cite{krisciunas:1991} was modified to better fit SQM observations (integrating the moonlight received within the field-of-view of the sensor) conducted under the typical atmospheric and light pollution conditions present in Hong Kong~\cite{so:2014}. The basic formula of this modified model (details at Supplementary Note 12) is: 
\begin{equation}
I_{\text{moon}}  = 2\int_{\phi  = 0}^{\phi _{\max } } {\int_{\theta  = 0}^{\pi} R(\phi) \cdot B_{\text{moon}} (\sin \phi) d\phi d\theta }, \label{eqt:I_moon2sqm}
\end{equation}
where 
$\phi_{\max}=20^{\circ}$ because the sensor's sensitivity drops to nearly zero at that angle~\cite{sqmlreport},
($\theta$, $\phi$) is the horizontal coordinate of the sky patch being integrated with the azimuth angle $\theta$ and zenith angle $\phi$, 
$R(\phi)$ is the sensor's azimuthally symmetric angular response function published~\cite{sqmlreport}.  

In 2013 during the Earth Hour event, the Moon was in a waxing gibbous phase (lunar phase = 0.84). The Moon's altitude above the horizon increased from 67$^\circ$ at 20:00, reached its highest point of 79$^\circ$ (meridian crossing) at 21:25 and then declined to 76$^\circ$ by 22:00. Given that the detection sensitivity of the employed SQM sensor is concentrated within a narrow cone of angle centered on the zenith, the amount of moonlight captured by the instrument increased gradually as the Moon approached the meridian in 2013 and then decreased gradually thereafter. The predicted contribution of moonlight to the observed NSB values ranged from 38\% to 51\%, depending on the altitude of the Moon.
In 2015, the altitude of the Moon in the first quarter (lunar phase = 0.62) decreased from 81$^\circ$ to 54$^\circ$ over the course of the two-hour period starting at 20:00. Consequently, the moonlight captured by the sensor is estimated to have gradually decreased from 34\% to 14\% as the Moon moved farther from the zenith.
In contrast, in 2021, the altitude of the near-full Moon (lunar phase = 0.98) increased from 37$^\circ$ to 63$^\circ$ during the two-hour window starting at 20:00. As a result, the moonlight contribution to the observed NSB is predicted to have gradually increased from 29\% to 40\% as the Moon approached the zenith.
The NSB light curves before and after the moonlight corrections for the years 2013, 2015 and 2021 are included in Supplementary figure S7.
The greatest predicted moonlight contribution during the Earth Hour period was 51\% when the Moon was at the meridian in 2013. However, the influence of the Earth Hour event is still discernible even before removing the lunar contribution for that time period.

It is important to note that the moonlight predictions presented here are only approximations, as the underlying model has an accuracy range of 8\% to 23\% ~\cite{krisciunas:1991}. Additionally, the model does not account for the shielding and scattering effects of clouds on moonlight. Consequently, the amount of moonlight actually received by the sensors may have been overestimated when the sky conditions were cloudy, such as between 20:00 and 20:27, and 21:34 and 22:00 during Earth Hour 2013, between 20:00 and 20:15 in 2015, and before 20:50 in 2021.

Unlike observations conducted at very dark locations~\cite{grauer:2019,duriscoe:2007}, the contributions from natural sources other than the Moon, such as stars, the Milky Way, zodiacal light, airglow and solar activity, are negligibly small compared to the contributions from ALAN at TST. Ref.~\cite{masana:2021} discusses contributions from natural sources.

\subsection*{Sky spectral observations}
The spectral data presented in this study was acquired using an \textsc{Ocean Insight} USB2000+ spectrometer at TST, same location as the all-sky and NSB observations. This handheld spectrometer features a 2,048-element CCD array detector that is sensitive to visible and near-infrared light in the wavelength range of 339.495 to 1027.681 nm. During the Earth Hour events, exposures were made continuously from 20:00 onward, with integration times of 65 seconds in 2021 and 17 seconds in 2022. To protect the spectrometer, it was placed inside a plastic storage container with a transparent cover. The spectrometer's aperture was pointed towards the zenith and calibrated with the standard Mercury-Argon lamp before and after the measurements. 

The standard data reduction procedure involved correcting each original spectrum for dark noise. The "before lights-out" and "after lights-out" spectra were then obtained by calculating the median of the spectra acquired between 20:00-20:25 and 20:45-21:15, respectively, for each year. Finally, the "before-and-after" comparison is visualized by plotting the normalized spectra and the residual in Figure~\ref{fig:spectra}.
To further verify that the observed residuals are attributed to the lights-out, we compared the sky spectra taken after the lights were turned back on (i.e., 21:35-22:00) with those taken during the lights-out period (i.e., 20:45-21:15). Similar kinds of residual spectral features were observed. Moreover, we could not detect any significant residuals from data taken on normal Saturdays when no coordinated lights-out event was occurring.

\bibliography{AAstyle}

\section*{Acknowledgements}
The authors acknowledge the Hong Kong Space Museum for providing all-sky images.
For Earth Hour 2021, the authors acknowledge WWF-Hong Kong and Airjason Li for providing footage, and the volunteers involved in the lights-out survey: Chin-Kwan Bartholomew Tsang, Chu-cheuk Jack Chi, Kang-chit Ambrose Wong, Chun-long Sunny Ho, Ka-chung Kan Chung, Ka-ho Ben Kwan, Kai-fung Ng, Tsz-chun Iu, Yin-cheung Marco Tung, Sum-yue Jenny Cheung.
Figure~\ref{fig:HKmap}: image courtesy of Image Science and Analysis Laboratory, NASA-Johnson Space Center, \textit{The Gateway to Astronaut Photography of Earth} (ISS042-E-152950); base map and data from \textit{OpenStreetMap} and \textit{OpenStreetMap Foundation}.
\textit{The Hong Kong Night Sky Brightness Monitoring Network} (NSN) was funded by the Environment and Conservation Fund (Project ID: 2009-10) of The Government of the Hong Kong Special Administrative Region. The \textit{Globe at Night - Sky Brightness Monitoring Network} (GaN-MN) is co-organized by the Office for Astronomy Outreach of the International Astronomy Union (IAU-OAO), The University of Hong Kong (HKU), National Astronomical Observatory of Japan (NAOJ) and Globe at Night. 
The Globe at Night citizen science campaign to monitor light pollution levels worldwide is hosted by the U.S. National Science Foundation National Optical-Infrared Astronomy Research Laboratory (NSF NOIRLab). GaN-MN was funded by the HKU Knowledge Exchange Fund granted by the University Grants Committee (Project No: KE-IP-2014/15-57, KE-IP-2015/16-54, KE-IP-2016/17-44, KE-IP-2017/18-54, KE-IP-2018/19-68, KE-IP-2019/20-54 and KE-IP-2020/21-78), the Environment and Conservation Fund (Project IDs: 125/2018, 113/2022) of The Government of the Hong Kong Special Administrative Region, IAU-OAO and NAOJ. 
The observations during Earth Hour 2013 were supported by the HKU Faculty of Science and student volunteers from the HKU Science Outreach Team.
Any opinions, findings, conclusions or recommendations expressed in this paper do not necessarily reflect the views of the Government of the Hong Kong Special Administrative Region and the Environment and Conservation Fund.

\section*{Author contributions}
C.W.S., C.S.J.P., S.L., S.L.C., H.K.K.H., K.B. and C.E.W. conceptualized the methodology and observations. 
C.W.S., C.S.J.P., S.L., S.L.C., H.K.K.H., K.B. and C.E.W. conducted the observations.
C.W.S., C.S.J.P. and S.L. analysed the data. 
C.W.S. and S.L. visualized the results.
C.W.S. wrote the main manuscript text. 
C.W.S., C.S.J.P., S.L. and C.E.W. edited the manuscript.
All authors reviewed the manuscript. 

\section*{Data availability} 
All raw data required for replication are available at \url{https://doi.org/10.6084/m9.figshare.28307198}

\section*{Supplementary Information} 
Supplementary information is available at \url{https://static-content.springer.com/esm/art%3A10.1038%2Fs41598-025-05279-4/MediaObjects/41598_2025_5279_MOESM1_ESM.pdf}

\end{document}